\documentclass[prl,
preprint,
amsmath,amssymb]{revtex4-1}

\usepackage{graphicx}
\usepackage{dcolumn}
\usepackage{bm}
\usepackage{units}
\usepackage{microtype}
\usepackage{hyperref}
\usepackage{upgreek}
\usepackage{booktabs}
\usepackage{longtable}

\begin{document}

\title{A permanent, stable, and simple top-contact for molecular electronics on Si: Pb evaporated on organic monolayers}
\author{Robert Lovrin\v{c}i\'{c}, Olga Kraynis, Rotem Har-Lavan, Abd-Elrazek Haj-Yahya, Wenjie Li, Ayelet Vilan, and David Cahen}
\email{david.cahen@weizmann.ac.il}
\affiliation{Department of Materials and Interfaces, Weizmann Institute of Science, Rehovot 76100, Israel}


\begin{abstract}
We show that thermally evaporated lead (Pb) preserves the electronic properties of organic monolayers on Si and the surface passivation of the Si surface itself. The obtained current-voltage characteristics are in accordance with results from the well-established hanging mercury drop method and preserve both the molecule-induced dipolar effect and length-attenuation of current. We rationalize our findings by the lack of interaction between the Pb and the Si substrate. Our method is fast, scalable, compatible to standard semiconductor processing, and can help to spur the large-scale utilization of silicon-organic hybrid electronics.
\end{abstract}

\maketitle


Top-contact fabrication on self-assembled organic monolayers (ML) is a critical issue in molecular electronics. If evaporated onto a ML with only  $\sim 1\,\mathrm{nm}$ thickness, the metal might penetrate through the flexible molecular array even in the absence of pinholes \cite{Silien08,Zhu06}, creating shorts that dominate the transport characteristics (\textit{diffusion} effect) \cite{Popoff12}.
Evaporated atoms may also react with the molecules, fully or partially destroying the electric properties of the ML (\textit{reaction} effect).
Last, the top-contact material might react with the substrate, which, for a semiconducting substrate, introduces interfaces states that could dominate the net junction behavior \citep{Har-lavan12} (\textit{pinning} effect).
A multitude of techniques has been developed over the years to contact organic MLs (cf.\ Ref.\ \cite{Haick08} for a review). Some of the most common include: a hanging drop of liquid Hg \cite{Yaffe10,Har-lavan12} (temporal), indirect evaporation of different metals \cite{Bonifas10,maitani_issues_2011,metzger_electrical_2001,popoff_preparation_2011,scott_fabrication_2007,wang_new_2010} (extremely slow), attaching pre-prepared conducting pads \cite{Vilan02}, transfer of CVD graphene on ML \cite{wang_new_2010} (ill-defined solvent trapping), spin coating of reduced graphene oxide \cite{seo_solution-processed_2012} and PEDOT/PSS \cite{park_flexible_2012} (limited conductivity of top-contact).

MLs on semiconductor (SC) surfaces can influence electron transport across a metal/ML/SC junction in two ways: (i) the molecular dipole changes the effective work function of the SC surface and, thereby, the height of the barrier in the SC for charge transport. (ii) The molecules form a tunneling barrier, the current decay through which depends on the molecular length. The relative importance of these two contributions depends on the energy levels of the molecules used, the doping and electron affinity of the SC, and the work function of the top-contact \cite{Vilan10}. The molecular dipole effect is mostly sensitive to reaction and pinning effects, while the molecular tunneling attenuation would be easily screened by a diffusion effect. 

Pb has been used as the standard top contact for inelastic tunneling spectroscopy (IETS) of organic layers on Al/Al$_2$O$_3$ since the 1960s \cite{Lambe68,Honciuc07}, by virtue of its low evaporation temperature and the superconducting transition at $T=7.2\,\mathrm{K}$.
The formation of Pb films \cite{Pucci06} and nanowires \cite{Klevenz08} on Si surfaces has been extensively studied because Pb does not chemically interact with Si and offers an excellent test system to investigate low dimensional electron transport in metals.
Moreover, nuclear-reaction analysis  indicates that, contrary to what is the case for Ag and Cu, hydrogen remains at the interface between Pb and the Si substrate even when Pb is evaporated on H-terminated Si at room temperature \cite{fukutani_hydrogen_1999,fukutani_nuclear-reaction_2001}. Also current-voltage characteristics on H-terminated n- and p-Si shows that the H-termination withstands Pb evaporation and the H-induced dipole influences transport characteristics \cite{kampen_lead_1995}.

This led us to the hypothesis that Pb could be a contact for organic MLs that will allow introducing the latter into Si electronics in a stable, permanent, and high-throughput manner. Note that the requirements for a good top-contact for $J-V$ measurements are more demanding than for IETS, as IETS is a lock-in technique that can tolerate if a part of the current does not go through the molecules. We will show in the following that Pb, evaporated on a variety of organic monolayers on Si (i) does not damage the head groups of two similar styrene MLs that differ only in their head group, -Br vs. -CH$_3$, based on the preservation of the molecule-induced dipole effect on electron transport across the Pb/Si junction and (ii) preserves a length dependence for tunneling through saturated alkyl chains of different lengths, suggesting that any pinhole formation that can short the molecules is negligible. For comparison, we evaporated Ti as a top contact, which clearly destroyed the molecular effect. We conclude that Pb can function as a permanent, stable, and easy to fabricate contact and can be used for molecular electronics with Si in general. Our results furthermore suggest that a lack of interaction between a vacuum-deposited contact and the substrate is crucial for preserving an interfacial ML intact.


Alkyl MLs were prepared in inert atmosphere using a Schlenk line \cite{Scheres10}, by thermal activation ($200\, ^{\circ}$C) on highly doped ($\rho=10^{-3}\, \mathrm{\Omega cm}$) Si(100) wafers, the methyl- and bromo-styrene ML were prepared by UV activation in an inert atmosphere glove box \cite{Firas12} on moderately doped ($\rho=10\, \mathrm{\Omega cm}$) Si(111), see references for details concerning ML formation. 
Before electrical characterization, the ML quality was checked with spectroscopic ellipsometry and water contact angle measurements. Additionally, we examined the ML quality by  $J-V$ measurements on the same samples with our standard top Hg-contact, before Pb (or Ti) deposition. Pb (or Ti)  was evaporated in a vacuum of $\sim 5\cdot 10^{-6}\, \mathrm{mbar}$ at a rate of $0.4\, \mathrm{nm/s}$ through a shadow mask to a thickness of $300\,\mathrm{nm}$ ($100\,\mathrm{nm}$).
Contact to the Si was made with an InGa-eutectic \cite{Firas12}. $J-V$ measurements were performed in a vacuum probe station (for Pb and Ti top-contacts) and a N$_2$ filled glove box (for Hg).


\begin{figure}
\includegraphics[width=8.5cm]{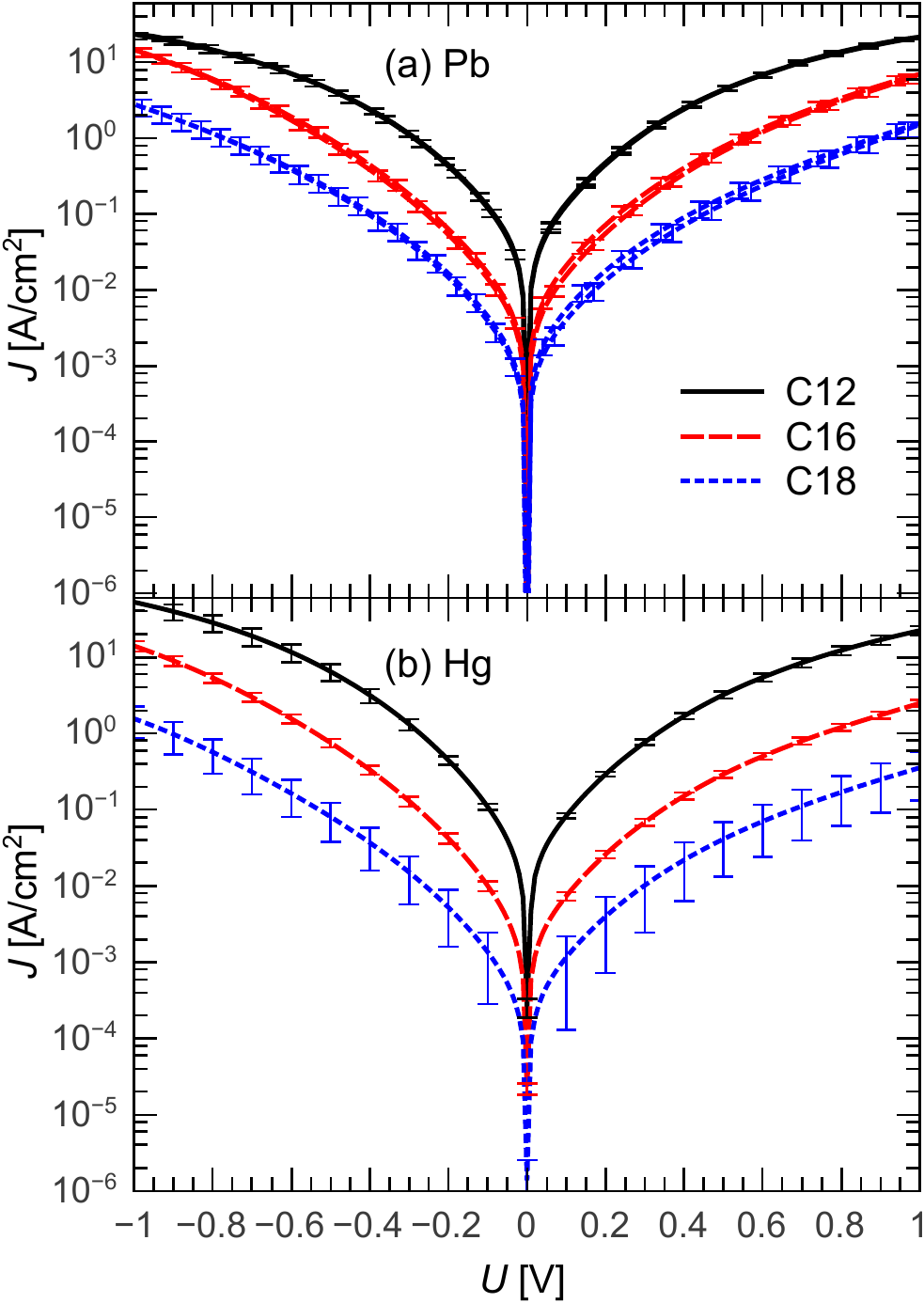}
\caption{\label{IVL}(color online).$J-V$ curves measured for alkyl chains of different lengths on highly doped n-Si(100) with (a) Pb and (b) Hg top contacts. Pb and Hg measurements were performed on the same samples. Contact diameter was $400\,\mathrm{\upmu m}$.}
\end{figure}

\begin{figure}
\includegraphics[width=8.5cm]{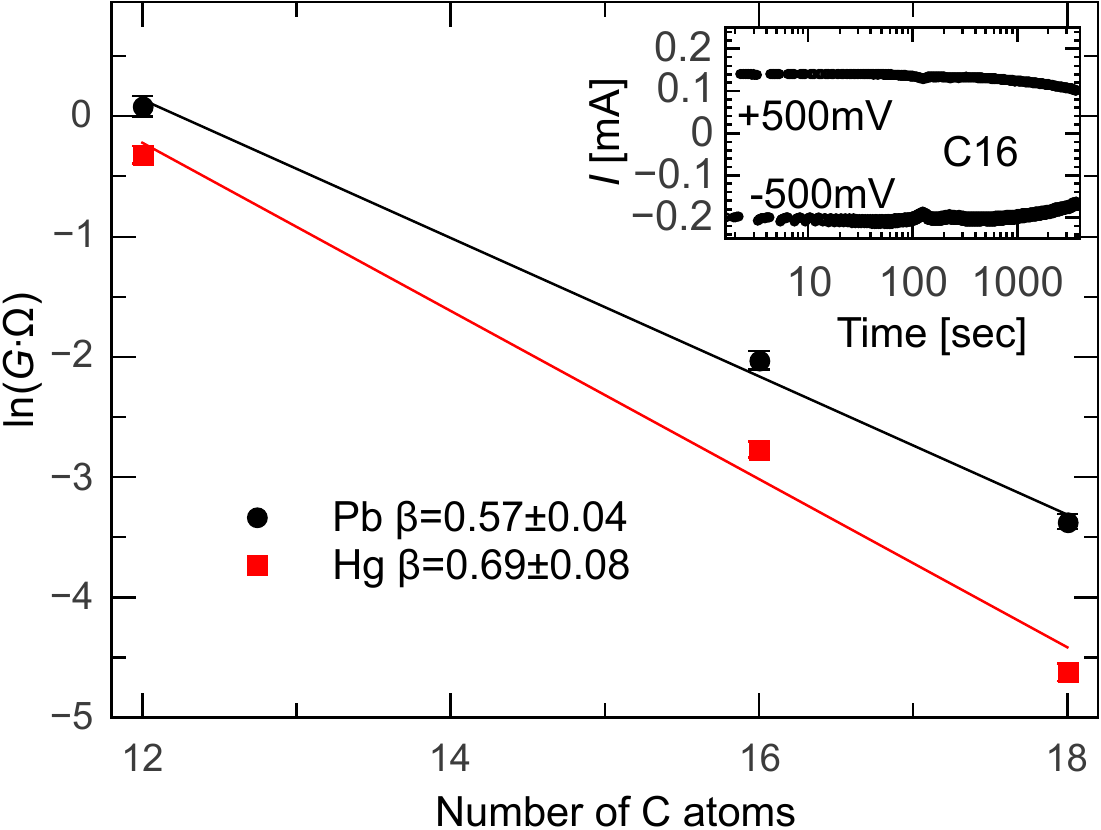}
\caption{\label{beta}(color online). Natural logarithm of the conductance vs. number of carbon atoms in the alkyl chains with both Pb (black circles) and Hg (red squares) top-contacts, extracted from the data shown in Fig.\ \ref{IVL} (for voltages $\leq 50\,\mathrm{mV}$). The tunneling decay parameter $\beta$ was extracted from the slope of the plots. Inset: retention characteristics for a C16 junction of the cross-measured positive (at $+0.5\,\mathrm{V}$) and negative (at $-0.5\,\mathrm{V}$) current with an interval of $\Delta t=1\, \mathrm{s} $.}
\end{figure}

Fig.\ \ref{IVL} shows $J-V$ curves, measured for alkyl chains of different lengths (12, 16, and 18 carbon atoms) on highly doped n-Si(100) with both Pb and Hg top-contacts. Each curve is an average over 7 different Pb pads and at least 3 different Hg drops measured on 2 samples per molecular length. Error bars are standard deviations. All graphs are almost symmetric with respect to bias polarity, as expected for highly doped Si, where the depletion layer is negligible. The behavior at high bias for C12 can be attributed to series resistance effects that become apparent at high currents. Note that these results were obtained on the same samples (Hg data measured before Pb evaporation). The data of all measured devices are shown in the supporting information. The inset in Fig.\ \ref{beta} shows  the cross-measured current through the C16 ML at $+0.5\,\mathrm{V}$ and $-0.5\,\mathrm{V}$ for one hour with a 1 sec measurement interval. Even under such repeated voltage stress conditions the current through the junction is stable within 20\%.

Charge carrier transport through saturated alkyl chains can be described as an elastic scattering problem, where the tunneling probability decays exponentially with molecular tunneling barrier length \cite{Yaffe10}. The conductance $G$ of a single channel can be written as 
\begin{equation}
G=G_{\mathrm{c}}\exp{(-\beta l)}.
\end{equation}
Here $l$ is the length of the molecule, $\beta$ is the length decay parameter, and $G_{\mathrm{c}}$ is the contact conductance. We extracted $\beta$ (per carbon atom) for the Pb and Hg junctions, as shown in Fig.\ \ref{beta}. To remain as close as possible to equilibrium and to avoid the influence of series resistance, we averaged the conductivity only in the limit of small bias ($U\leq 50\, \mathrm{mV}$). The $\beta$ values for Pb and Hg agree within error bars ($0.57\pm 0.04$ for Pb and $0.69\pm 0.08$ for Hg). A $\beta$ value around 0.6 per C falls within the lower limit of reported values for alkyl chains \cite{Wang12,Yaffe10}. However, as the Hg contacts gave similar results, we attribute this to the MLs on n-Si(100), rather than to the Pb top-contact. The pronounced exponential length-attenuation of the current indicates that the Pb contact does not cause pinholes through the ML (at least not more than Hg) because pinholes conduct orders of magnitude more than the ML and thus would dominate the net current even at very low pinhole densities. The good reproducibility of the measured data is an additional, indirect evidence for the lack of pinholes, which are random (see for comparison the behavior with Ti top contact, SI).

Temperature dependent $J-V$ measurements for Pb/C16/n-Si (highly doped) showed that, as expected for electron tunneling through alkyl chains \cite{Wang03}, transport is not thermally activated (see supporting information). This is a further indication that transport is not dominated by pinholes, for which case a thermal activation would be expected, due to the Si barrier.

\begin{figure}
\includegraphics[width=8.5cm]{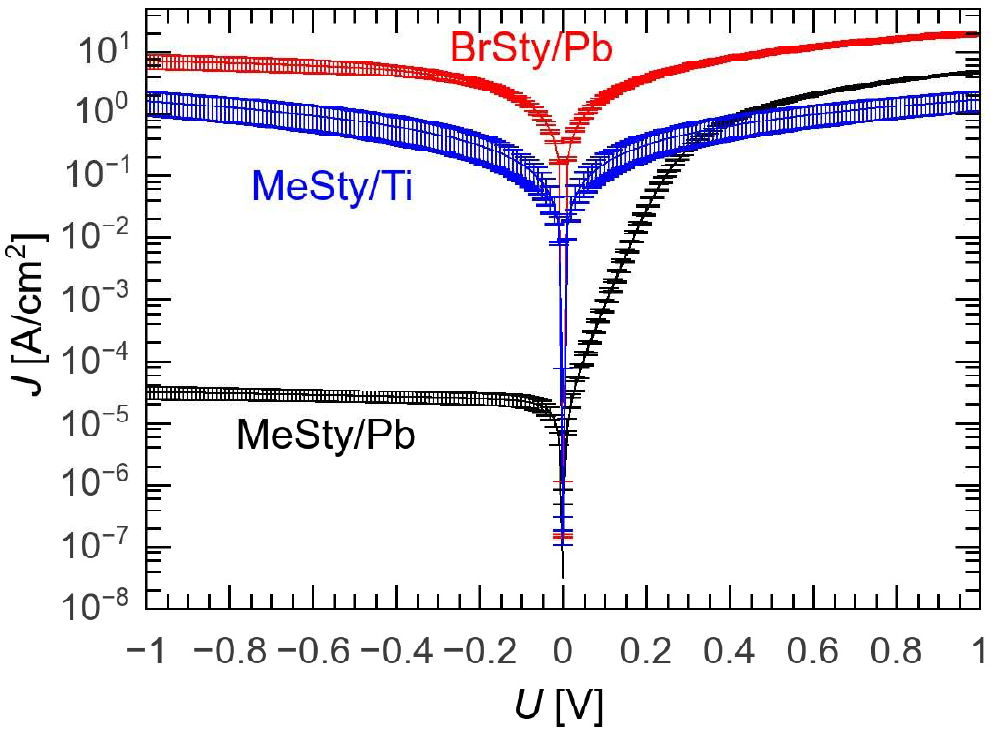}
\caption{\label{dipole}(color online). $J-V$ curves measured for methyl- and bromo-styrene on moderately doped Si(111) with Pb top-contact and Ti for comparison. Contact diameter was $400\,\mathrm{\upmu m}$ . See text for barrier heights.}
\end{figure}

To check that Pb does not react with an organic monolayer and preserves the molecular dipole effect, we compared two MLs of X-styrene-derivatives molecules that differ only in the head group, X: bromo- and methyl-styrene. We showed that with Hg contacts (Hg/X-styrene/Si(111) junction) the different dipole induced by the substituent X leads to extremely different $J-V$ curves. A Hg/bromo-styrene/Si(111) junction shows ohmic behavior, whereas  Hg/methyl-styrene/Si(111) is highly rectifying with a Schottky barrier height $\phi_{\mathrm{SBH}}$ of $0.81\,\mathrm{eV}$ \cite{Firas12}. This can be understood considering the different electron affinities of the modified Si surfaces ($3.6\,\mathrm{eV}$ for methyl-styrene and $4.4\,\mathrm{eV}$ for  bromo-styrene \cite{Firas12}).
According to the Schottky-Mott model, for an n-type semiconductor with electron affinity $\chi$ and a metal with
work function $\varphi_{\mathrm{m}}$ the energy barrier at the interface $\phi_{\mathrm{SBH}}$ is defined as \cite{Ng07}:
\begin{equation}
\phi_{\mathrm{SBH}}=\varphi_{\mathrm{m}}-\chi .
\end{equation}
$\phi_{\mathrm{SBH}}$ can be extracted from a $J-V$ curve using the thermionic emission equation \cite{Ng07}:
\begin{equation}
\phi_{\mathrm{SBH}}=\frac{kT}{q}\ln \left( \frac{AT^2}{J_0}\right) .
\end{equation}
$A$ is the Richardson constant ($\sim 100\,\mathrm{Acm^{-2}K^{-2}}$) and $J_0$ the saturation current (taken at $U=-0.2\,\mathrm{V}$).

Fig.\ \ref{dipole} shows the $J-V$ curves for Pb/  methyl- or bromo-styrene on moderately doped Si(111) and, as comparison, Ti on  methyl-styrene. Ti was used for comparison because it has the same work function as Pb ($\varphi_{\mathrm{Pb}}= 4.2\,\mathrm{eV}$ \cite{Michaelson77}, Hg has a work function of $\sim 4.5\,\mathrm{eV}$) but, in contrast to Pb and Hg, does react with Si. Moreover, it has been reported that Ti forms a good contact on MLs with a CH$_3$ top group \cite{Zhou97,Chang03}. As can be seen, Pb forms an almost ohmic junction with the bromo-styrene ML, and a highly rectifying with methyl-styrene, in accordance with the Hg results \cite{Firas12}. The $0.65\,\mathrm{eV}$ extracted $\phi_{\mathrm{SBH}}$ for Pb on methyl-styrene corresponds well with the barrier height for Hg on methyl-styrene, as the difference can be mostly attributed to the difference in metal work function and the uncertainties in the metal work function values  ($\varphi_{\mathrm{Hg}}- \varphi_{\mathrm{Pb}}=0.25\,\mathrm{eV}$ \cite{Michaelson77}, $\phi_{\mathrm{SBH\ Hg}}-\phi_{\mathrm{SBH\ Pb}}=0.16\,\mathrm{eV}$).
This implies that the Si surface remains well passivated and the Pb evaporation does not induce surface states that would cause Fermi level pinning \cite{Har-lavan12}.

Ti on methyl-styrene, on the other hand, shows ohmic behavior, meaning that the molecular dipole effect is completely destroyed upon Ti evaporation. To check that this is not an effect of surface heating during the evaporation, but rather the result of a chemical reaction of the Ti with the molecules and substrate, we measured $J-V$ curves on the same sample with Hg in an area not covered with Ti. Indeed the junction was highly rectifying (see SI).

\begin{figure}
\includegraphics[width=8.5cm]{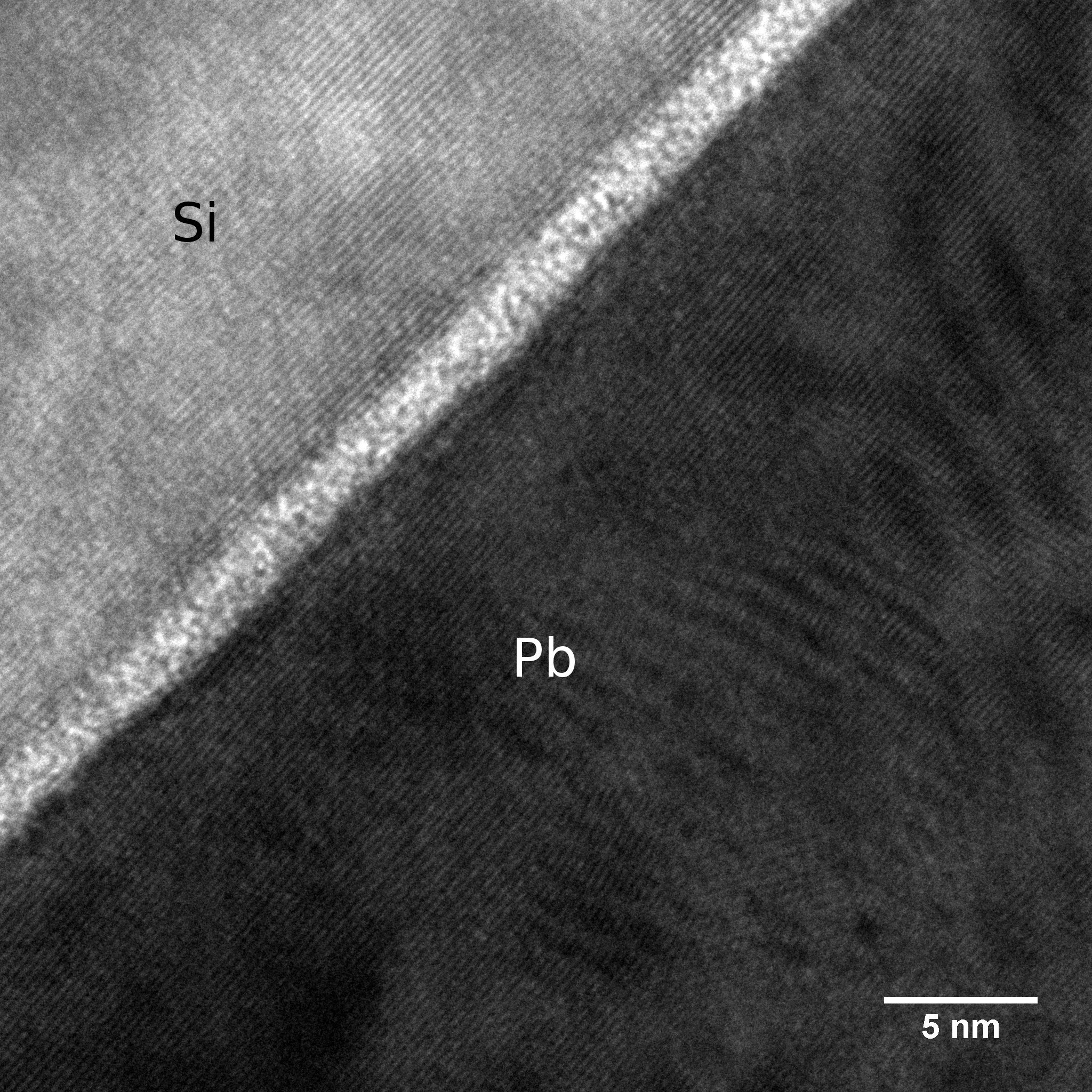}
\caption{\label{TEM}(color online). Cross section TEM measurement for a Si(111)/Br-styrene/Pb junction.}
\end{figure}

Further proof for the non-interacting nature of the directly evaporated Pb top-contact is the transmission electron microscopy measurement shown in Fig.\ \ref{TEM}. The sample was Pb/bromo-styrene/Si(111). No direct Pb/Si contact was observed; instead, there is a uniform $\sim 2\,\mathrm{nm}$ interlayer with a low mass density.	Such a clean interlayer has previously been reported only for Au, evaporated on a ML with a thiol group at the top that reacts with the Au and prevents it from penetrating \cite{Kuikka08}. Apparently such a chemical barrier is not necessary for our Pb contacts, making them a very versatile top-contact for molecular electronics on Si.


Beyond the specific conclusion that evaporated Pb forms a non-destructive, reliable permanent contact to monolayers on Si, our findings are important for the general considerations in making contacts to any organic monolayers (or, in general, nm thin layers of soft matter). While it is clear now that metal penetration can occur at perfect monolayer regions due to molecular flexibility and is not limited to pre-deposition defects \cite{Silien08,Zhu06}, it is generally accepted that a strong affinity between molecular head-groups and deposited metal is the main mechanism that blocks pinholes formation \cite{Kuikka08}. However, various works pointed to the substrate and its tendency to react with the deposited atoms as the major player dictating the destruction of the delicate monolayer. This was shown for evaporation of different metals onto alkyl-thiol based monolayers on a Au substrate \cite{Haynie03}, electrochemical deposition of Cu on monolayers of biphenyl thiols on Au \cite{Silien08}, and evaporation of different metals on alkyl monolayers on Si and SiO$_2$ \cite{Hacker07}. Our results lend strong credence to the importance of also this second mechanism. At the same time, the comparison between Ti and Pb evaporated contacts also shows that too strong interaction with a head group is at best problematic for taking advantage of the electrostatic effects of dipolar molecules for controlling semiconductor junctions. Finally, the Pb/X-styrene-Si(111) junctions are the first demonstration of standard, solid-state junctions controlled by a molecular dipolar effect, with a fabrication procedure which is readily available for technology implementation.


RL is grateful to the European Union for a Marie Curie postdoctoral fellowship (grant FP7-PEOPLE-2011-IEF no. 298664) and to the Minerva Foundation (Munich) for a fellowship during the first part of this work. We thank the Israel Science foundation via a Center of excellence and the Grand Center for Sensors and Security for partial support. DC holds the Schaefer chair in Energy research.

\providecommand*{\mcitethebibliography}{\thebibliography}
\csname @ifundefined\endcsname{endmcitethebibliography}
{\let\endmcitethebibliography\endthebibliography}{}

\end{document}